\documentclass[preprint2]{aastex}

\newcommand{\simgt}{\,\rlap{\lower 3.5 pt \hbox{$\mathchar \sim$}} \raise
1pt \hbox {$>$}\,}
\newcommand{\simlt}{\,\rlap{\lower 3.5 pt \hbox{$\mathchar \sim$}} \raise
1pt \hbox {$<$}\,}

\shorttitle{Cores and density profiles}
\shortauthors{Hansen \& Stadel}

\begin{document}


\title{The origin of cores and density profiles 
of gaseous baryonic structures}


\author{Steen H. Hansen \& Joachim Stadel}
\affil{University of Zurich, Winterthurerstrasse 190,
8057 Zurich, Switzerland}



\begin{abstract}
  We study the origin of cores and density profiles of gaseous baryonic
  structures in cosmology.  By treating the baryons as a viscous gas,
  we find that both spheres and disks are possible solutions.  We find
  analytically that the density profiles have inner and outer
  solutions, which in general are different. For disks we identify a
  central core, with density profile $\rho_d = constant$, and the
  outer profile $\rho_d \sim r^{-3}$. For spherical structures we find
  the profile $\rho_s \sim r^{-6}$.  In the presence of a dominating
  central black hole we find the inner profile $\rho \sim r^{-3/2}$.
  When the mass is
  dominated by a dark matter component then the baryonic density
  profile will depend on the dark matter profile, and we point out
  how one can use this connection to infer the DM profile directly by
  observing the baryonic density profile. 
\end{abstract}




\section{Introduction}

The universe is full of large gaseous baryonic structures such as galaxies and
clusters. These structures are observed to take on a variety of shapes
ranging from disks to spherical configurations. The radial profiles of
these structures have been observed to be rather non-trivial,
e.g. changing from one power-law profile near the centre to another
power-law profile at large radii. Even though these baryonic
configurations have been observed and studied for many years, there is
surprisingly little theoretical guidance to understand the origin of
such complexity. Astronomers have for years been using
phenomenological profiles,
ones with differing inner and outer shape,
however, no simple
theoretical explanation for this structure exists.  

In recent years numerical analyses have improved enormously, and we
can now simulate much of the structures we observe. 
Never the less, it is
important to have a simple understanding of the
underlying physics, which can be obtained most easily through
analytical studies of the basic equations.

In this letter we 
attempt an analytical treatment of
the basic equations, asking which density profiles are expected of
purely gaseous baryonic structures, and which profiles lead to stable
baryonic structures within dark matter halos.  We address these
questions by considering a fluid approach whereby we analyse
asymptotic stable solutions of the Navier-Stokes equations.  This
approach is new in astrophysics and surprisingly simple.  
Our main results are, that the density profile of gaseous baryonic 
structures is fairly
complex, namely that the radial density profile may have different
slopes in the inner and outer regions.
This is an issue which has been much discussed for Dark
Matter profiles, but which has not been understood previously for
baryonic structures.  We find that spheres and disks are the only
possible solutions.  From an observational point of
view it seems obvious that these are possible solutions, and while
the creation of disks is understood theoretically,
the fact that no other solutions are stable is non-trivial.  As a
byproduct of our analysis we show how one can infer the DM density
profile purely by observing the baryonic density profile.  We also
analyse how a central massive black hole influences the solutions.

\section{Solving Navier-Stokes equations}

The behaviour of any collisional gas or fluid is fully determined by
the Navier-Stokes (N-S) equations, which are 3 hydrodynamical
equations for the velocity vector, and a continuity equation.  
In the
following we will make several assumptions, which certainly limits the
applicability of our results, however, the solutions we find should be
valid for the description of general properties of
structures such as gaseous baryonic galaxies and haloes.  The main assumption
is that there are sufficient collisions while the structure forms.
Thus if a gaseous baryonic structure has formed and subsequently all the
baryons condense into stars, then the star density profile should
still approximately follow the original density profile.  
Baryons often have sufficient collisions to
equilibrate, e.g. in a typical intra-cluster gas the 
equilibration timescale is
about $10^7$ yrs, with mean free path of tens of kpc compared to
radii of few Mpc.
We do not expect that
our findings should apply to dark matter structures, since dark matter
presumably does not have sufficient collisions to ensure the validity of
the N-S equations.

In spherical coordinates the N-S equations describe the velocity vector
$(v_r, v_\Theta, v_\phi)$. Here
$r$ is the radial coordinate, $\Theta$ is the angle in the $xy$-plane,
and $\phi$ is the angle from the $z$-axis. The form of the equations
is well known, see e.g.~\citet{ll87}, and contains time-derivatives,
pressure terms, viscosity terms and a gravitational term.

Our first 2 assumption are, that the gas has reached a stable
configuration (no time-derivative), and that it has picked out an
orientation in space, in such a way that all the gas is moving only in
the $\Theta$-direction. Thus we have $v_r =0$ (no contraction or
expansion), and $v_\phi=0$.
Here one must keep in mind, that by
considering the N-S equations we are taking a fluid approach, which
implies that we are following a fluid element, and this basically
corresponds to averaging over all the particles moving through the
fluid element. 
For the
$\Theta$-velocity we consider the general form
\begin{eqnarray}
v_\Theta &=& v_\alpha \, \left( \frac{r}{r_\alpha}  \right)^\alpha \,
\left( {\rm sin}\phi \right) ^\chi \, , \label{eq:v}
\end{eqnarray}
where $\alpha$ and $\chi$ are constants to be determined, $v_\alpha$
and $r_\alpha$ are unknown constants, with the physical interpretation
that $r_\alpha$ is a characteristic radius,
and $v_\alpha$ is the velocity of the fluid element at that radius.

\subsection{The $\Theta$-equation}

The $v_\Theta$-equation becomes very simple with the assumed form of
the velocities
\begin{equation}
0 = \nu \, \left[ \nabla^2 v_\Theta  
- \frac{v_\Theta}{r^2 {\rm sin}^2 \phi} \right] \, ,
\label{eq:vis}
\end{equation}
where $\nabla^2$ is the scalar Laplacian, and $\nu$ is the kinematic
viscosity. For now all that matters is the existence of a non-zero
viscosity, so the absolute magnitude (and even radial dependence) is
not important for the results~\footnote{Certainly baryons have a
  non-zero viscosity, however the viscosity could also arise from
  turbulence in which case it could be very large, $\nu_{\mbox{turb}}
  \sim l \Delta v$~\citep{ll87}, where the dimension $l$ is the size of
  the turbulent eddies, and $\Delta v$ is the velocity dispersion.}.
When we use the form for $v_\Theta$ in eq.~(\ref{eq:v}), then
equation~(\ref{eq:vis}) has 4 solutions
\begin{eqnarray}
\chi = +1 &,& \alpha = 1,-2 \, , \label{eq:plus}\\
\chi = -1 &,& \alpha = 0,-1 \, . \label{eq:minus}
\end{eqnarray}

The solution with $\chi = +1$ is exactly what one should expect for
spherical symmetry, and we will refer to these solutions as
the {\em 'spherical solutions'}. The solutions with $\chi = -1$
indicate unstable solutions everywhere except in the flat cylinder
where ${\rm sin} \phi =1$, and we will refer to these as {\em 'disk
  solutions'}.  We thus see that both spherical and disk solutions
exist, and that they have different rotational structure (different
$\alpha$).  Now, looking at eq.~(\ref{eq:v}), it is clear, that the
solutions with negative $\alpha$ are divergent for $r \rightarrow 0$,
and we will therefore refer to those solutions as {\em 'outer
  solutions'}, and similarly, the solutions with positive $\alpha$ are
inconsistent for large radii, and we will refer to those as {\em
  'inner solutions'}.


There are several obvious extensions of the method described here.
One could in particular allow for a general $\Theta$-dependence
in $v_\Theta$, a non-zero radial 
velocity, or an exponential radial dependence of velocities and
density profile~\footnote{Allowing a $\Theta$ dependence
in $v_\Theta$ as $v_\Theta \sim (\mbox{sin} n\Theta)^\sigma$,
with $n$ the number of spiral arms, gives $\sigma=0,1$ for all $n$.
Allowing for a non-zero radial velocity, $v_r \sim r^\tau$, gives 
$\tau = 1,-2$.}.

Already from this one sees that a spherical solution, in
eq.~(\ref{eq:plus}), must have different radial structure for the
inner and outer solutions (because the only solutions are $\alpha = 1,
-2$), and hence one may expect to find different density profiles in
the central and outer regions. We point out that such
phenomenon of simultaneous existence of two flow patterns is rather
common in hydrodynamics, the simplest may be the hydraulic jump which
is observed as a several centimetre large circular ring in any kitchen
sink, when the flowing water goes from a $1/r$ profile to a constant,
see \citet{hansen97,bohr}.
We emphasize that a non-zero viscosity appears to be a
necessary condition for the existence of these specific solutions, even
though a fundamental understanding of how the microscopic physics
(viscosity) can determine the macroscopic properties (the general flow
pattern) is still generally missing in hydrodynamics.

\subsection{The $r$-equation}

From the $\Theta$-equation, we identified the general flow pattern,
and we noticed the possibility that one may have  
different
velocity-flows
in the inner and outer region.  
We will now use the $v_r$-equation to try to extract
the asymptotic radial density profiles. Also the $r$-equation is very
simple
\begin{equation}
-\frac{v_\Theta^2}{r} = -\frac{1}{\rho}\frac{\partial P}{\partial r}
- \frac{M(r) G}{r^2} \label{eq:vr} \, ,
\end{equation}
where $\rho$ is the radially dependent density, $P$ is the pressure,
$G$ is the gravitational constant, and $M(r)$ is the mass within the
radius $r$. 
We assume that the pressure and density are related
through $P = P_\alpha\, (\rho/\rho_\alpha)^\gamma$, where $P_\alpha$
and $\rho_\alpha$ are the unknown pressure and density at $r_\alpha$.
We assume the gas is monatomic with $\gamma=5/3$.  Let us consider
densities of the form
\begin{equation}
\rho(r) = \rho_\alpha \, \left( \frac{r}{r_\alpha}\right)^\beta \, ,
\label{eq:rho}
\end{equation}
such that the parameter $\beta$ determines the density profile.  It is
worth emphasising that it is exactly this $\beta$ which we are trying
to find.

Let us study the radial dependence of the 3 terms in
eq.~(\ref{eq:vr}).  Using $v_\Theta$ in
eq.~(\ref{eq:v}) the first (kinetic) term of eq.~(\ref{eq:vr}) goes
like $v_\theta^2/r \sim r^{2\alpha-1}$.  The pressure gradient term
goes like $1/\rho\cdot \partial P/\partial r \sim r^{\delta \beta-1}$,
where we have used $\delta = \gamma -1 = 2/3$. The last
(gravitational) term including $M(r)$,
depends on the given system we are considering. If the mass is
dominated by a point gravitational source (e.g. a central black hole), 
then it goes like $M(r) G/r^2 \sim
r^{-2}$.  If the mass is dominated by the matter density, then it goes
like $M(r) \sim \int \rho(r) dV$, with $dV$ the volume element.  For
spherical solutions this gravitational term thus goes like
$r^{\beta+1}$ with $\beta$ from eq.~(\ref{eq:rho}), and for disk
solutions it goes like $r^\beta$.  
This gravitational term has the correct form for spherical distributions 
(and point sources), but is only an approximation for the pure
disk case.
Technically speaking the mass is
logarithmic divergent for spherical structures with $\beta=-3$,
however, the formula $M \sim r^{\beta + 3}$ holds for any $\beta$
arbitrarily close to $-3$, and furthermore in a real situation there
would be an outer cut-off.
To be explicit, we are looking for solutions to an equation
of the form
\begin{equation}
r^{2\alpha -1} = r^{\delta \beta-1} + r^\kappa \, ,
\label{eq:simpel}
\end{equation}
where $\kappa = \beta+1, \beta, -2$ for spherical, disk and BH matter
dominance respectively.  When we use the word 'solve' in the following, 
we are really just using the standard method of divergence cancellation, 
in the sense that the most
divergent terms must cancel with each other. The optimal case is
naturally that all divergences disappear, a case which we  will
refer to as {\em 'good'}.  From our simple analysis the
transition radius, $r_\alpha$, which separates the inner from outer
region, is not uniquely determined.  We only find approximate disk
relations like $r_\alpha \sim v^2_\alpha/G \rho_\alpha$.  A
full study including the coefficient is significantly more involved,
and we will leave that for a later analysis~\footnote{We note 
that when the $v_\Theta^2$ term is negligible
this eq.~(\ref{eq:vr}) is exactly the hydrostatic equilibrium, which is often
written as $M(r) \sim rT \gamma \beta$. Furthermore, in the same
notation one finds that the velocity dispersion, $\sigma_v^2$, 
has a radial dependence,
$ d {\rm ln} \sigma_v^2 /d {\rm ln} r = \delta \beta$.}.

We want to solve eq.~(\ref{eq:vr}) for the profile parameter $\beta$,
however, there are 4 situations to consider (the 4 different $\alpha$
from eqs.~(\ref{eq:plus},\ref{eq:minus})), and for each case we can
choose which is the dominating mass contribution, either from a
sphere, from a disk or from a black hole (BH).
The discussion in sections 2.3 and 2.4 is somewhat technical, and the
reader is encouraged to look at figs.~1 and 2, which summarize the
main findings of these sections.

\subsection{Disk}
\label{sec:disk}

Let us first consider the $\alpha=0$ case, which can be both an inner and
outer solution. For the inner solution we find
two density profiles, $\beta =0,-3$. The $\beta=0$
solution is more natural, because the first two divergent terms in
eq.~(\ref{eq:vr}) (or similarly eq.~(\ref{eq:simpel})) can cancel, and
the last term is non-divergent for $r \rightarrow 0$, whereas for the
$\beta = -3$ solution, the two most divergent terms cancel, but the
remaining term is still slightly divergent towards the centre. With
this kind of argument one can divide all the solutions into {\em
  'good'}, and {\em 'reasonable'} solutions, and we will
emphasise which are {\em good} solutions.  If the mass is dominated by
a BH, then the solution is $\beta =-3/2$. If the disk is dominated by
a spherical distribution (either from a large fraction of the baryons,
or from an unknown dark matter component) with profile $\beta_s$, then
the solution is $\beta = 3/2\,(\beta_s + 2)$.  In conclusion we see,
that the only good solution for the inner disk is $\beta=0$.
In principle the $\alpha=0$ case can also be an outer solution. In
that case the solution is $\beta = -1$. If the mass is dominated
by a BH, then $\beta =0$ (good), and if dominated by a spherical
distribution then $\beta \leq 0$.


\begin{figure}[htb!]
\plotone{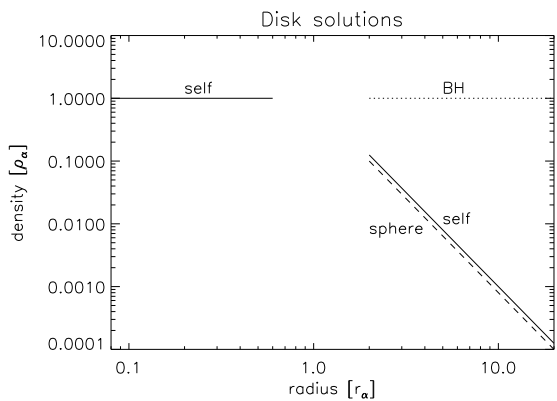}
\caption{A summary of disk solutions. 
The full lines ({\em self}, $\beta = 0,-3$) are
where the mass is dominated by the disk itself. The dotted line ({\em
BH}, $\beta=0$) is for black hole domination, and the dashed line
({\em sphere}, $\beta=-3$) is for matter domination by a sphere with
profile , $\beta_s=-3$.}
\end{figure}

The $\alpha = -1$ (which is an outer solution) gives $\beta=-3$ (good). 
If a BH dominates then
$\beta=-3/2$, and if another spherical component dominates (with
profile $\beta_s$) then $\beta = 3/2 \, (\beta_s + 2)$, where the case
$\beta=-3$ is the only good solution.
We summarize the
good disk solutions in Figure 1.

To conclude the disk solutions, when the mass is dominated by the
matter itself, then there is only 1 good configuration (we emphasise
that by good we mean optimal removal of divergences), which has the
inner slope of $\beta=0$ and outer slope of $\beta=-3$.  Following the
similarity with the hydraulic jump in the kitchen sink, we will refer
to such configuration as a galactic hydraulic drop.  

\subsection{Sphere}
\label{sec:sphere}
For the sphere it turns out that almost all the solutions are good,
in the sense that all the divergences cancel in a simple manner. 
The $\alpha=1$ is the inner solution, for which  we find $\beta = 3,-6$.
The $\beta = 3$ is obviously strange, and probably non-physical
(a positive $\beta$ would lead to a wrong sign
in front of the pressure gradient term). 
With BH dominance one has
$\beta = -3/2$, and if dominated by a disk (which could either be from 
a fraction of the same gas, or from another particle type) with profile
$\beta_d$ one finds $\beta = 3/2 \, (\beta_d + 1)$. If dominated by 
another spherical distribution (which probably should arise 
from another particle type) with profile $\tilde \beta_s$, then we find
$\beta = 3/2 \, (\tilde \beta_s + 2)$.

The outer solution, $\alpha = -2$ gives $\beta = -6$. If a BH
dominates then $\beta = -3/2$, and again if a disk (or sphere) dominates then
$\beta = 3/2 \, (\beta_d + 1)$ (or $3/2 \, (\tilde \beta_s + 2))$.
We summarize the good
spherical solutions in Figure 2.

\begin{figure}[htb!]
\plotone{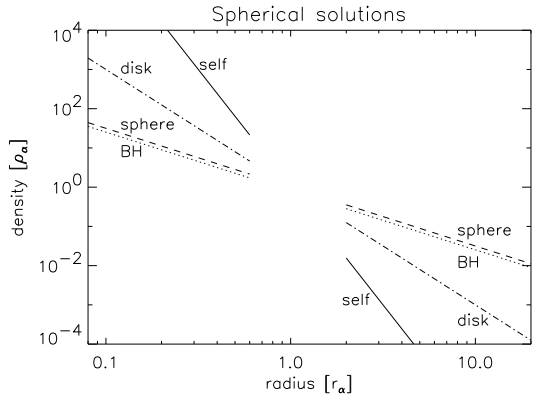}
\caption{A summary of spherical solutions. 
The full lines ({\em self}, $\beta=-6$) are where the mass is
  dominated by the sphere itself, and the dotted lines ({\em BH}, $\beta=-3/2$) are
  for black hole domination.  The dashed lines ({\em sphere}, $\beta=-3/2$) are for
  matter domination by another sphere with profile -3, and the
  dot-dashed lines ({\em disk}, $\beta=-3$) are for domination of a disk with
  profile -3.}
\end{figure}


We note that if the mass is disk dominated with
disk-profile $\beta_d = -3$, then the sphere happens to also get the
profile $\beta_s = -3$. Thus one can envisage a significant fraction
of the matter collapsing to a disk, which very well could take the
profile $\beta_d=-3$ (as shown in section~\ref{sec:disk}), and this
would force the remaining matter, which is in a spherical
configuration, to take the density profile $\beta_s = -3$.  We also
point out that if the mass is dominated by a dark matter sphere, then
the baryon profile becomes $\beta=-3/2$ for a dark matter profile
$\tilde \beta_s = -3$ as suggested by dark matter N-body
simulations. If instead the dominating DM profile is $\tilde \beta _s
>-2$, this would seemingly imply a positive baryonic slope, which is
difficult to interpret.  
We feel that this limited applicability warns, that 
a more general calculation may provide a different 
connection between the DM and gaseous baryonic profiles

\section{Discussion}

An interesting possibility now appears, namely that one can use our
results to infer the dark matter profiles from observations of the
baryonic profile. 
This method can be used quite generally to infer the DM distribution,
and is therefore complementary to other methods such as lensing
observations.
Let us say we have observed a baryonic sphere with
profile $\beta_{\rm baryon}$, and we know that the mass is
dark matter dominated. Under the assumption that the DM is spherical
we have $\beta_{\rm baryon} = 3/2
\, (\tilde \beta_{DM} + 2)$. Thus, if we observe e.g.  $\beta_{\rm
baryon} = -1.5$, then we know, that the DM has profile $\tilde
\beta_{DM}= -3$. Determining the baryonic density profile directly, i.e. 
independent of dynamics, can be accomplished in several ways:

1) X-rays: observations of the luminosity, $L_{x}$, in various bands
and in different radial bins gives the radial electron density of the 
plasma. Here the main concern is the validity of hydrostatic
equilibrium and disentangling any cooling flow in the centre of the
cluster. Relaxed clusters with no evidence of cooling flows
do exist, for example A2029 (see Figure~1 of \citep{lewis03}). The gas
in this cluster shows an inner profile $\beta_{\rm baryon} = -0.55$
and an outer $\beta_{\rm baryon} = -1.62$ (note our differing
definition of $\beta$ here). The outer value is certainly DM dominated
and our analysis implies that $\tilde\beta_{DM}$ is very close to
$-3$, which is expected from CDM simulations \citep{nfw96,moore99}.
We do not attempt to deduce the inner DM slope here, because one
cannot be certain that the mass is DM dominated at such small radii
where the baryonic slope should reach its asymptotic value.
Again we refer to the warning in the end of section~\ref{sec:sphere}.

2) The Sunyaev-Zeldovich effect is in principle a direct measure of
the plasma column density. The angular resolution required for our
analysis currently limits the use of this technique, however, future
multi-frequency observations will determine independently both the
temperature and number density profile of distant clusters purely
through the S-Z effect~\citep{hansen02,aghanim03}. The SZ-effect can thus be
used to measure the electron density profile to large radii (SZ effect
is proportional to $n_e$, whereas X-ray is proportional to $n_e^2$)
and at large redshift (SZ is redshift independent).

3) Surface brightness: both from stellar light and radio observations
of HI and molecular gas, one can in principle determine the baryon
profile. An example is
for M33 by \citep{corb03}. 

The distribution near black holes has been considered earlier.
First by~\cite{peebles72} where energy consideration lead to the
distribution $\rho \sim r^{-9/4}$, which was refined
in~\citep{bahcall76} who found $\rho \sim r^{-7/4}$.
Later numerical simulations have shown \citep{young80}
that the profile near the black hole should be $\rho \sim r^{-3/2}$,
which is just what we find. For accreting black holes this profile may be
different \citep{freitag02}.

\section{Conclusion}

We make a first attempt to derive analytically the asymptotic density
profiles of baryonic structure in cosmology, which include galaxies,
gaseous haloes and intra-clusters gas. 
We find that both disks and spherical solutions
exist, and that generally the inner and outer density profile may be
different.  
Thus we supply theoretical support for the use of phenomenological
profiles like
\begin{equation}
\rho_{\rm gas} (r) = \frac{\rho(0)}{r^{\beta_1} 
( 1 + r )^{\beta_2}}  \, .
\end{equation}
For the disks we identify central cores with $\beta_d=0$
from $\rho \sim r^\beta$. For the outer region we find $\beta_d=-3$.
For spherical structures we identify both inner and outer profiles,
which include $\beta_s = -6, -3, -3/2$. 
Our resulting profiles only apply to gaseous baryonic structures, but
we point out a simple method whereby observations of the baryonic
structure in principle allows one to deduce the dark matter density profile.

\acknowledgments
It is a pleasure to thank Tomas Bohr,
James Binney, Sasha Dolgov, Lucio Mayer, Ben Moore and the people in Zurich
for useful discussions,
Victor Debattista and Marcella Carollo for observational
guidance,
and in particular Ruth Durrer for a valuable discussion.



\end{document}